\documentstyle[twocolumn,prl,aps,epsf]{revtex}
\begin{document}

\draft
\wideabs{
\title{Effect of Charge Fluctuations on the  Persistent Current 
 through a Quantum Dot}
\author{Kicheon Kang$^{1,2}$, Sam Young Cho$^3$, Ju-Jin Kim$^2$}
\address{   $^1$Basic Research Laboratory, Electronics and Telecommunications
            Research Institute, Taejon 305-350, Korea}
\address{   $^2$Department of Physics, Chonbuk National University,
            Chonju 561-756, Chonbuk, Korea}
\address{   $^3$Institute of Physics and Applied Physics, Yonsei University,
            Seoul 120-749, Korea}

\date{\today}

\maketitle

\begin{abstract}
 We study coherent charge transfer between an Aharonov-Bohm ring
 and a side-attached quantum dot.
 The charge fluctuation between the two sub-structures
 is shown to give rise to algebraic suppression of the persistent current
 circulating the ring 
 as the size of the ring becomes relatively large.
 The charge fluctuation at resonance
 provides transition between the diamagnetic and the paramagnetic states.
 Universal scaling, 
 crossover behavior of the persistent current from a continuous to a 
 discrete energy limit in the ring is also discussed.
\end{abstract}
}
%
%

 Coherent electron transfer from one region into another region in a composite
 mesoscopic structure may affect drastically the quantum 
 mechanical properties of the coupled system. 
 Coupling between the subsystems due to the electron transfer
 accompanies charge and/or spin fluctuation, and modifies significantly
 the characteristics of the system.
 A prototype to investigate such kind of effects is
 a mesoscopic Aharonov-Bohm (AB) ring tunnel- coupled to a quantum 
 dot (QD)~\cite{buttiker96,ferrari99,kang00,eckle00,cho01,affleck01,hu01,anda01}
 or connected to a finite-size wire
 ~\cite{buttiker94,deo95,anda97,pascaud97,cedraschi98,cho98}.
 A purely one-dimensional (1D) ring of spinless electrons exhibits
  persistent current (PC) being either diamagnetic or paramagnetic depending
 on the number of electrons being odd or even~\cite{leggett91}, respectively.
 This so called ``Leggett's conjecture" breaks down when the 
 ring is coupled to 
 an additive structure such as a QD, 
 because the system is not a pure 1D any longer. 
 In a side-branched QD case of Ref.~\cite{buttiker96},
 it has been shown that coherent charge transfer
 from a QD to the ring induces sharp transition between
 plateaus of diamagnetic and paramagnetic states.
 This phenomenon is quite understandable 
 in terms of single charge transfer that alters 
 the number of electrons in
 the ring and in the QD one by one. 
 Their discussion, however, 
 was limited to a relatively small ring
 where the energy level spacing ($\delta$) of the ring 
 is much larger than the tunnel coupling strength ($\Gamma'$) 
 between the QD and the ring. 

 In the meantime, the role of spin fluctuation 
 associated with the Kondo effect on the PC 
 has recently become an issue of serious 
 debates~\cite{kang00,eckle00,cho01,affleck01,hu01}.
 The central point of the debates is whether the following argument of
 the correspondence between an open system and its closed counterpart
 can be applied to the system with a Kondo correlated quantum dot:
 {\em The PC in a 1D non-interacting
 scattering model with a fixed electron number $N_e$
 is completely determined by the transmission
 probability at the Fermi level, to the leading order of $1/L$ with
 $L$ being the length of the ring}~\cite{gogolin94}.   
 Transport current (TC) through a 1D quantum wire with a side-branch QD 
 is completely suppressed in the Kondo limit~\cite{kang01}. 
 On the contrary,
 it has been shown by the exact Bethe ansatz~\cite{eckle00} and by 
 a diagrammatic expansion~\cite{cho01}
 that the PC of a mesoscopic ring with a side-branch QD is 
 not affected by the presence of the Kondo effect 
 in the continuum limit (supporting the spin-charge separation).
 On the other hand, Affleck and Simon~\cite{affleck01}
 have obtained the opposite result of Ref.~\cite{eckle00,cho01}
 by using a renormalization group (RG) 
 analysis combined with the above described argument of Ref.~\cite{gogolin94},
 and concluded that the normalized PC is completely suppressed in the 
 continuum limit.

 At this stage, there arise the fundamental questions
 associated with strong charge fluctuations till now not clarified:
 (i) whether the PC and the TC 
 are straightforwardly interrelated each other or not, and
 (ii) how is the universal scaling behavior of the PC in the limit of
 $\delta/\Gamma'\rightarrow 0$.
%
 To clarify these points let us consider a QD 
 side-coupled to a mesoscopic ring.
 To understand the effect of charge fluctuation on the PC and the TC,
 it will be very instructive to study the case of the noninteracting QD
 because this model is exactly solvable.
 We show that charge fluctuations between the ring and the QD 
 suppress the PC algebraically.
 We argue that the suppression can be understood without the assumption that 
 the PC is a Fermi surface effect.
 Although this result cannot be directly applied
 to the system with the Kondo correlation,
 the comparison between the PCs due to the charge fluctuation
 and due to the spin fluctuation may provide deeper insight
 to resolve the ongoing issue of the Kondo-correlated quantum dot
 side-coupled to a mesoscopic ring.

 We begin with the model Hamiltonian of the system:
 \begin{mathletters}
 \begin{equation}
  H = H_0 + H_D + H_T ,
 \end{equation}
 where $H_0$, $H_D$, and $H_T$ represent the AB ring,
 the QD, and tunnel coupling, respectively. 
 \begin{eqnarray}
 H_0 &=& -t \sum_{j=0}^{N-1} \left( e^{i\phi/N} c_j^\dagger c_{j+1} +
     \mbox{\rm h.c.} \right) \;, \\ 
 H_D &=& \varepsilon_d d^\dagger d \;, \\
 H_T &=& -t' \left( d^\dagger c_0 + c_0^\dagger d \right).
\end{eqnarray}
\label{eq:hamil}
\end{mathletters}
 For the ring we employ 
 a 1D periodic tight-binding model
 of $N$ lattice sites ($c_N=c_0$) with the hopping integral $t$ taken
 to be real without loss of generality. The phase factor $\phi$ 
 is defined by $\phi=2\pi\Phi/\Phi_0$  with
 $\Phi$ and $\Phi_0=hc/e$ being the AB flux and 
 the flux quantum, respectively. 
 The QD, which is much smaller than the ring, 
 is modeled as a single resonant level of the energy $\varepsilon_d$. 
 For our concern, 
 this model is sufficient to study
 the effect of a charge fluctuation on the PC. 
 The QD is coupled to the ring at site `0' 
 by tunneling matrix element $t'$.
 Throughout our discussion,  
 $t'$ is considered to be much smaller than $t$.
 Also, for simplicity, we consider only half-filled case, 
 $N_e=(N+1)/2$ with odd $N$,
 where the total number of lattice sites including the dot is $N+1$.
 This assumption does not affect the conclusion 
 we will draw in the present study.
%

 The essential parameter representing the coupling strength between the
 ring and the QD, namely $\Gamma'$, is defined by
 \begin{equation}
 \Gamma' = \pi \rho(\varepsilon_F) |t'(\varepsilon_F)|^2 ,
 \end{equation}
 where $\rho(\varepsilon_F)$ and $t'(\varepsilon_F)$
 represent the density of states and the hopping amplitude
 at the Fermi energy $\varepsilon_F$, respectively.
 For half-filled case ($\varepsilon_F=0$) at continuum limit,
 $\rho(0)=\delta^{-1}=N/(2\pi t)$ and $t'(0)=-t'/\sqrt{N}$.
 Thus, $\Gamma'$ is given by 
 \begin{equation}
  \Gamma' = \frac{t'^2}{2t} .
 \end{equation}

 The Hamiltonian (\ref{eq:hamil}) can be exactly diagonalized, 
 which gives the $N+1$ eigenvalues $\{E_\alpha\}$. 
 After some algebra, 
 one finds that the eigenvalues can be obtained from the equation
 \begin{equation}
 E_\alpha - \varepsilon_d 
  = \frac{\Gamma'}{\pi}\delta \sum_m \frac{1}{E_\alpha -\varepsilon_m}, 
 \label{eq:eigen}
 \end{equation}
 where $\varepsilon_m =  -2t\cos[(2m\pi-\phi)/N]$ 
 are the eigenvalues of the ring Hamiltonian $H_0$ 
 with $m$ being integers corresponding to the angular momenta. 

 At zero temperature, 
 the PC $I(\phi)$ is given by the
 coherent charge response of the ground state to the AB flux as
 \begin{equation}
 I(\phi) = -\frac{e}{\hbar} \sum_\alpha^{occ} 
            \frac{\partial E_\alpha}{\partial\phi} ,
 \label{eq:pc}
 \end{equation}
 with the summation to be taken only for occupied levels of $\{E_\alpha\}$.
 Combining Eq.(\ref{eq:eigen}) and (\ref{eq:pc}) one get the expression
 for the current
 \begin{mathletters}
 \begin{equation}
  I(\phi) = \sum_m F(\varepsilon_m) I_m(\phi) , 
 \end{equation}
 where 
 \begin{equation}
 I_m(\phi) = -\frac{e}{\hbar}\frac{\partial\varepsilon_m}{\partial\phi} 
 \end{equation}
 is the current contribution by the 
 bare energy level $\varepsilon_m$ for the ring, and
 \begin{equation}
   F(\varepsilon_m) = \frac{\Gamma'}{\pi} \delta
    \sum^{occ}_\alpha \frac{A_\alpha}{ (E_\alpha -\varepsilon_m)^2 }
 \end{equation}
 represents the effective distribution function of electrons at the
 level $\varepsilon_m$ with
 \begin{equation}
  A_\alpha = \left[ 1 + \frac{\Gamma'}{\pi}\delta
    \sum_m\frac{1}{(E_\alpha-\varepsilon_m)^2} \right]^{-1} .
 \end{equation}
 \end{mathletters}
 $F(\varepsilon_m)$ corresponds to the probability of an electron
 occupied in the bare energy level $\varepsilon_m$,
 with its summation on $m$ being just the average number of electrons
 in the ring part:
 \begin{equation}
  \sum_m F(\varepsilon_m) = N_e - n_d ,
 \end{equation}
 where $n_d$ is the occupation number of electrons in the QD. 

\begin{figure} 
\epsfxsize=3.0in
\epsffile{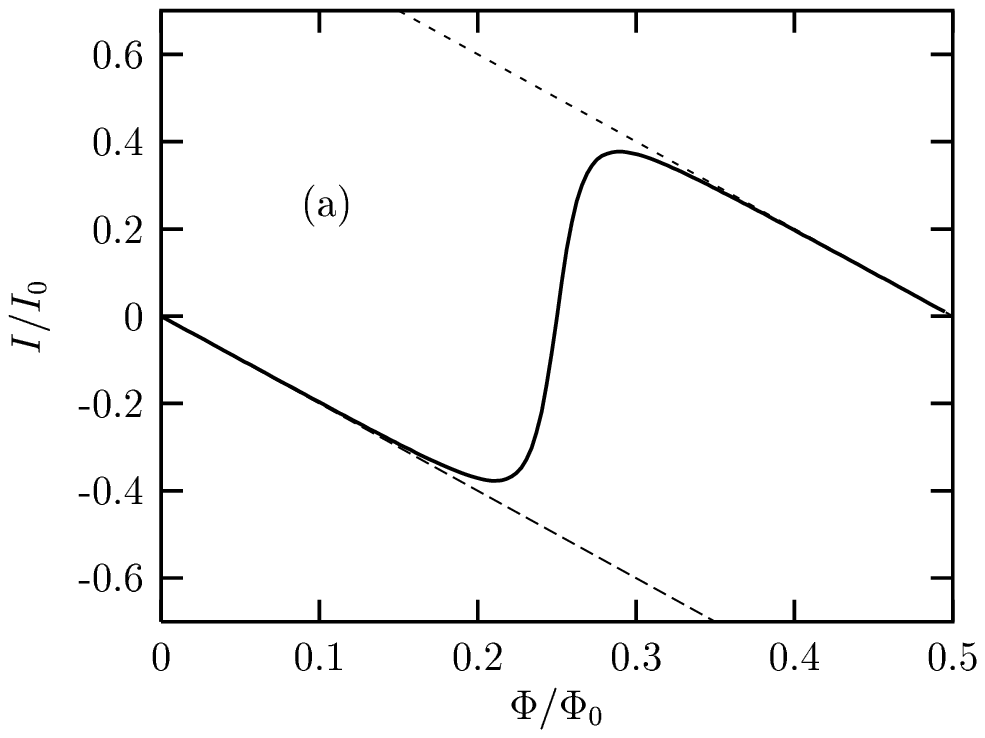}
\epsfxsize=3.0in
\epsffile{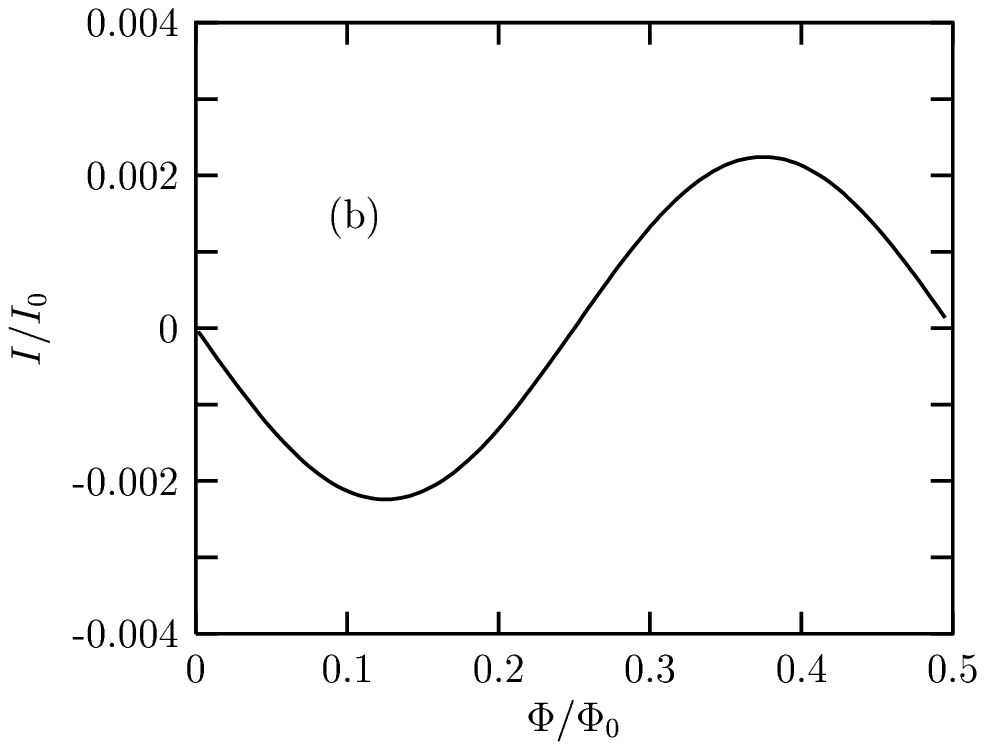}
 \caption{ Charge resonant ($\varepsilon_d=0$) persistent current 
  - phase relation (a) for weak coupling
  ($\delta/\Gamma'=1000$), and (b) for strong coupling ($\delta/\Gamma'
  =0.5$). The dashed and the dotted lines in (a) correspond to the current
   of an ideal ring with the number of electrons being odd and even,
   respectively.  }
  \label{fig1}
\end{figure}

 First, 
 we discuss the current-phase relation
 at resonance ($\varepsilon_d=0$) 
 where the charge fluctuation 
 is strongest.
 Fig.~\ref{fig1} shows the PC in two different cases - (a) weak
 and (b) strong coupling, respectively. 
 $I_0$ denotes the PC amplitude of an ideal ring with the same radius,
 $I_0=ev_F/L$ with $v_F$ being the Fermi velocity. 
%
 Some interesting results are found both in the weak and in 
 the strong coupling cases.
 First, the current displays no parity effect, that is, it does not depend on
 the number of electrons being even or odd in contrast to pure 1D systems.
 Second, its period is half ($\Phi_0/2$) of that in an isolated ring. 
 It should be noted that $I/I_0$ is a universal function of 
 $\delta/\Gamma'$ for $\varepsilon_d=0$.
%

 For the level spacing $\delta$ much larger than the coupling strength
 $\Gamma'$ (Fig. \ref{fig1}(a)), 
 the behavior of the PC shows a crossover from 
 diamagnetic to paramagnetic state of an ideal ring with 
 a sawtooth-shaped functional form. 
 The crossover of the ground state occurs at the AB flux $\Phi=\Phi_0/4$, 
 which leads to the period halving of the PC.
 This crossover can be understood in the following way. 
 Only the topmost level of the ring with $N_e$-electrons 
 mainly contributes to the PC
 in the $\delta\gg\Gamma'$ limit (to be precise, in the $\delta\gg t'$ limit). 
%
%
 The PC follows that of an ideal ring with $N_e$ electrons 
 or with $N_e-1$ electrons, 
 depending on whether the topmost level is occupied or not.
 The criterion, which curve the PC will follow,
 depends on the relative magnitude of $E_{N_e}^0$ and $E_{N_e-1}^0$ 
 where $E_{N_e}^0$ represents the ground state energy of an ideal ring
 with $N_e$-electrons.  
 With this criterion, one can find that the ground
 state of the system should be diamagnetic
 at $\Phi<\Phi_0/4$, and paramagnetic at $\Phi>\Phi_0/4$. 
 This fact is not affected 
 by the total number of electrons $N_e$ being even or odd,
 and thus the parity effect disappears in this special case.

 Fig.~\ref{fig1}(b) shows that,
 for a relatively strong coupling
 $I/I_0$ is much suppressed due to the strong charge fluctuation
 and level hybridization. 
 The current-phase relation resembles a sinusoidal function.
 This behavior is quite universal in the $\delta/\Gamma'\rightarrow0$ limit
 with a much suppressed value of $I/I_0$.
 Here we emphasize that {\em this feature of the transition from a 
 diamagnetic to paramagnetic state is a unique phenomenon of charge
 fluctuation which is not present in a Kondo system with $\delta
 \rightarrow0$ limit}. 
 That is, the effect of charge fluctuation does not provide one-to-one
 correspondence to that of the spin fluctuation studied in
 Refs. \onlinecite{eckle00,cho01,affleck01}.

\begin{figure}
\epsfxsize=3.0in
\epsffile{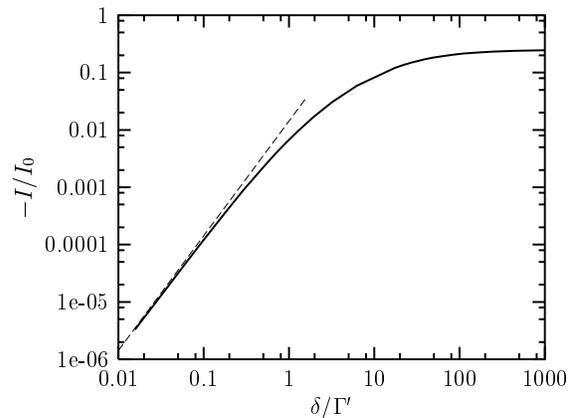}
\caption{ Universal curve of $I/I_0$ on resonance ($\varepsilon_d=0$)
 as a function of $\delta/\Gamma'$, for $\Phi=\Phi_0/8$.
 Dashed line corresponds to $I/I_0 \propto -(\delta/\Gamma')^2$.}
 \label{fig2}
\end{figure}

 The universal feature of the PC as a function of 
 $\delta/\Gamma'$ is well understood in Fig. \ref{fig2}.
 $I/I_0$ as a function of $\delta/\Gamma'$ is plotted in log-log scale.
 $\Phi=\Phi_0/8$ is chosen for this figure but the overall feature
 including the universal scaling is independent of a given value of the flux. 
 In the very weak tunneling limit ($\delta/\Gamma' \rightarrow\infty$),
 the current saturates to a value of an ideal ring, 
 $I/I_0=-0.25$ for $\Phi=\Phi_0/8$.
 This is exactly what is expected from Fig. \ref{fig1}(a). 
 $I/I_0$ is diminished as $\delta/\Gamma'$ decreases.
 One can find that at
 small $\delta/\Gamma' (\ll 1)$ the curve of the PC
 becomes linear with its slope being exactly two. 
 This implies that $I/I_0$ is proportional to $(\delta/\Gamma')^2$ 
 in this region, and eventually reduces to zero
 in the continuum limit of $\delta/\Gamma' \rightarrow 0$
 ($L \rightarrow \infty$).

 The vanishing of $I/I_0$ at $\delta/\Gamma' \rightarrow 0$ limit 
 could be interpreted in two different ways as follows. 
 First interpretation is based on an
 analogy between the open system and the ring with 
 $L\rightarrow\infty$~\cite{gogolin94}.
 It was shown in Ref.~\cite{kang01} 
 that the linear response
 conductance through a 1D perfect quantum wire with a side-coupled QD
 is given by
 \begin{equation}
 G = \frac{2e^2}{h} T(\varepsilon_F) 
 \end{equation}
 where $T(\varepsilon_F)=\cos^2\pi n_d$ is the transmission
 probability at the Fermi level for spinless electrons. 
 At resonance, $n_d=1/2$, thus the conductance reduces to zero. 
 This perfect reflection is understood as a result of 
 the destructive interference 
 between the direct transmission through the wire 
 and the resonant transmission via the QD.
 The vanishing of the normalized PC in our study might be regarded
 as the same characteristics with the perfect reflection in the corresponding
 open system.

\begin{figure} 
\epsfxsize=3.0in
\epsffile{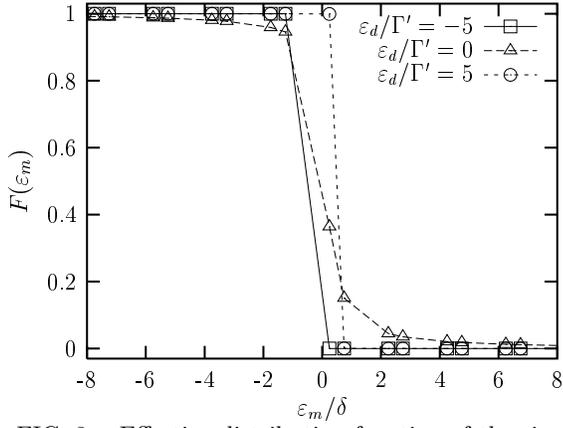}
\caption{ Effective distribution function of the ring energy levels in 
 the strong coupling limit ($\delta\ll\Gamma'$). The parameters are given as
 $N_e=388$, $\Phi=\Phi_0/8$, and $\delta/\Gamma'=5.06\times 10^{-2}$. 
 }
 \label{fig3}
\end{figure}

 However, the following alternative viewpoint
 can elucidate the suppression of the PC in a closed system without introducing 
 the analogy with the open system.
 For $\varepsilon_d=0$ with $\delta/\Gamma'\ll 1$, various configurations
 of distribution in the ring states contribute to the current,
 with its total number of electrons being $N_e$
 or $N_e-1$. These include many excited states with its excitation
 energy lower than $\Gamma'$ with the number fluctuation of
 electrons in the ring part. Each configuration contributes to the
 current with its magnitude and size depending very much on the 
 configuration of excited states.
 These configurations modify the occupation weight of each states
 $\varepsilon_m$ in the ring.
 The effective distribution is plotted in Fig. \ref{fig3}
 as a function of the ring energy level.
%
%
 One can find that, 
 for $\varepsilon_d\ll -\Gamma'$, 
 $F(\varepsilon_m)$ coincides with
 the distribution function of an ideal ring with $N_e-1$ electrons.
 That is, $F(\varepsilon_m) = 1$ for $N_e-1$ lowest energy states, and
 $F(\varepsilon_m) = 0$ otherwise. 
 Similarly, for $\varepsilon_d \gg \Gamma'$, 
 $F(\varepsilon_m)$ corresponds to the distribution function of 
 an ideal ring with $N_e$-electrons. 
 On the other hand, at charge resonance point ($\varepsilon_d=0$), 
 $F(\varepsilon_m)$ shows partial occupation weight
 for the levels adjacent to $\varepsilon_d$.
 Actually, this induces the algebraic suppression of the PC.
 It is important to note that this alternative interpretation does not
 require an assumption that the PC can be regarded
 as a Fermi surface effect.
 Furthermore, this argument cannot be applied to the Kondo system
 where the charge fluctuation is completely suppressed.

\begin{figure}
\epsfxsize=3.0in
\epsffile{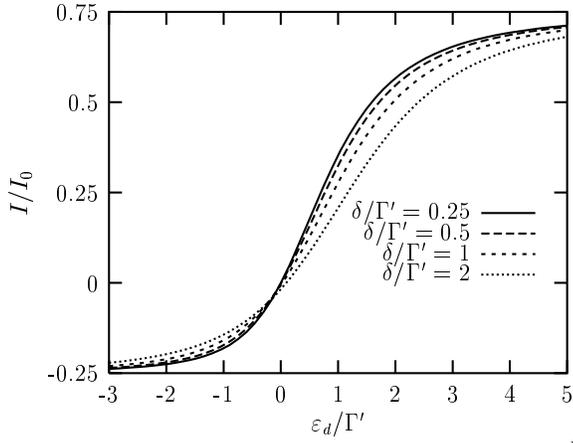}
\caption{ Crossover of the persistent current from $I_{ideal}^{N_e-1}(\phi)$
 to  $I_{ideal}^{N_e}(\phi)$ as a function of the dot level position. 
 Here the parameters are given as $N_e=188$, $\Phi=\Phi_0/8$. 
 }
 \label{fig4}
\end{figure}

 In Fig. \ref{fig4},
 the PC as a function of $\varepsilon_d$
 is displayed for different values of $\delta/\Gamma'$.
 The current shows a crossover from $I=I_{ideal}^{N_e-1}$ at $\varepsilon_d
 \ll -\Gamma'$ to $I=I_{ideal}^{N_e}$ at $\varepsilon_d\gg\Gamma'$.
 The crossover occurs around 
 $-\varepsilon_d^*\leq\varepsilon_d\leq\varepsilon_d^*$ where 
 $\varepsilon_d^*\simeq\Gamma'$, for small
 enough $\delta/\Gamma'$. In fact, $I/I_0$ is expected to be a universal 
 function of the renormalized parameter $\varepsilon_d/\Gamma'$ in the limit of
 $\delta/\Gamma'\rightarrow0$. 
 One can find 
 that the crossover value of the dot level 
 $\varepsilon_d^*$ increases as the level discreteness of the ring
 becomes important. 
 In the $\delta/\Gamma'\gg1$ limit, the crossover value $\varepsilon_d^*$
 does not correspond to $\Gamma'$ any more. This can be easily seen
 by considering the extremely discrete limit of the ring where
 $\delta\gg 2t'$. In this limit, only the topmost energy level, namely
 $\varepsilon_L$, will participate
 in the charge transfer to the dot~\cite{buttiker96}. 
 Then one can write the current in the following simplified form
 \begin{equation}
  I(\phi) = I_{ideal}^{N_e-1}(\phi) 
   - \frac{e}{2\hbar}\frac{\partial\varepsilon_L}{\partial\phi}
   \left( 1 + \frac{\varepsilon_d-\varepsilon_L}{ \sqrt{(\varepsilon_d
           -\varepsilon_L)^2 + (2t')^2 } } 
  \right) \, ,
 \end{equation}
 from which one can find that $I\simeq I_{ideal}^{N_e-1}$ at 
 $\varepsilon_d-\varepsilon_L\ll -2t'$ and $I\simeq I_{ideal}^{N_e}$ at
 $\varepsilon_d-\varepsilon_L\gg 2t'$. The crossover energy value is 
 $\varepsilon_d^*\simeq 2t'$ instead of $\Gamma'$. Because $\Gamma'=t'^2/2t
 \ll 2t'$ in our study $\varepsilon_d^*$ increases as $\delta$ increases,
 which indicates that the discreteness of the ring eigenstates weakens
 the effect of charge transfer resonance.
 A similar kind of evolution from the continuum to the discrete limit
 was studied in Ref.~\cite{sandstrom97} for
 a double-barrier resonant tunneling model, which shows a transition
 from ``true resonance" to ``semi-resonance" behavior of the PC. 
 Our case can be considered as another manifestation of the 
 crossover from the ``true resonance" to the `semi-resonance"
 for the case of a side-attached QD.

 In conclusion, the effects of resonant charge 
 transfer between a QD and an AB ring
 have been investigated by considering the PC circulating the ring.
 We have discovered a nontrivial algebraic suppression of the PC
 due to the charge fluctuations in the continuum limit of ring energy levels. 
 We have also found that 
 the transition of diamagnetic to paramagnetic states is closely
 interrelated to the PC suppression, and
 that this property does not provide an exact one-to-one correspondence
 to the coherent charge response to the AB flux at Kondo resonance.
%



%
%

\begin{references}
%
\bibitem{buttiker96} 
 M. B\"uttiker, and C. A. Stafford, \prl {\bf 76}, 495 (1996).
\bibitem{ferrari99} V. Ferrari, G. Chiappe, E. V. Anda, and M. A. Davidovich,
 Phys. Rev. Lett. {\bf 82}, 5088 (1999).
\bibitem{kang00} K. Kang and S.-C. Shin, Phys. Rev. Lett. {\bf 85}, 5619 
 (2000); J. Korean Phys. Soc. {\bf 37}, 145 (2000).
\bibitem{eckle00} H.-P. Eckle, H. Johannesson, and C. A. Stafford,
 J. Low Temp. Phys. {\bf 118}, 475 (2000); Phys. Rev. Lett. {\bf 87},
 016602 (2001).
\bibitem{cho01} S. Y. Cho, K. Kang, C. K. Kim, and C.-M. Ryu,
 Phys. Rev. B {\bf 64} 033314 (2001). (cond-mat/0011215)
\bibitem{affleck01} I. Affleck and P. Simon, \prl {\bf 86}, 2854 (2001);
 \prb {\bf 64}, 085308 (2001).
\bibitem{hu01} H. Hu, G.-M. Zhang, and L. Yu, \prl {\bf 86}, 5558 (2001).
\bibitem{anda01} E. V. Anda, C. Busser, G. Chiappe, and M. A. Davidovich,
 cond-mat/0106055 (2001).
\bibitem{buttiker94}
 M. B\"uttiker, Phys. Scr. {\bf T54}, 104 (1994).
\bibitem{deo95}
 P. Singha Deo, \prb {\bf 51}, 5441 (1995).
\bibitem{anda97}
 E. V. Anda, V. Ferrari, and G. Chiappe, J. Phys. Condens. Matter
  {\bf 9}, 1095 (1997).
\bibitem{pascaud97}
 M. Pascaud and G. Montambaux, Europhys. Lett., {\bf 37}, 347 (1997).
\bibitem{cedraschi98}
 P. Cedraschi, and M. B\"uttiker, J. Phys. Condens. Matter {\bf 10}, 
  3985 (1998).
\bibitem{cho98}
 C.-M. Ryu and S. Y. Cho, \prb {\bf 58}, 3572 (1998).
\bibitem{leggett91} A. J. Leggett, in {\em Granular Nanoelectronics},
 edited by D. K. Ferry {\em et al.}, NATO ASI Ser.
 B, Vol. 251 (Plenum, New York, 1991), p. 297.
\bibitem{gogolin94} A. O. Gogolin and N. V. Prokof'ev, \prb {\bf 50}, 
 4921 (1994).
\bibitem{kang01} K. Kang, S. Y. Cho, J.-J. Kim, and S.-C. Shin, \prb {\bf 63},
 113304 (2001).
\bibitem{sandstrom97} P. Sandstr\"om and I. V. Krive, \prb {\bf 56}, 9225
 (1997).
%
\end{references}
\end{document}